\documentstyle[preprint,tighten,aps]{revtex}
\begin{document}
\preprint{cond-mat/9310010}
\title{Universality of Br\'{e}zin and Zee's spectral correlator}
\author{C.W.J. Beenakker}
\address{Instituut-Lorentz, University of Leiden\\
P.O. Box 9506, 2300 RA Leiden, The Netherlands}
\date{October 1993}
\maketitle
\begin{abstract}
The smoothed correlation function for the eigenvalues of large hermitian
matrices, derived recently by Br\'ezin and Zee [Nucl.\ Phys.\ B402 (1993) 613],
is generalized to all random-matrix ensembles of Wigner-Dyson type.
\end{abstract}
\narrowtext
\section{Introduction}
A basic problem in random-matrix theory is to compute the correlation of the
eigenvalue density at two points in the spectrum, from the Wigner-Dyson
probability distribution of the eigenvalues \cite{Meh67}. The correlation is a
manifestation of the level repulsion resulting from the jacobian
$\prod_{i<j}|\lambda_{i}-\lambda_{j}|^{\beta}$, which is associated with the
transformation from the space of $N\times N$ hermitian matrices to the smaller
space of $N$ eigenvalues $\lambda_{1},\lambda_{2},\ldots\lambda_{N}$. [The
power $\beta$ depends on whether the matrix elements are real ($\beta=1$,
orthogonal ensemble), complex ($\beta=2$, unitary ensemble), or quaternion real
($\beta=4$, symplectic ensemble).] The Wigner-Dyson probability distribution
\begin{mathletters}
\label{PWdef}
\begin{equation}
P(\{\lambda_{n}\})=Z^{-1}\,\exp[-\beta W(\{\lambda_{n}\})],
\label{Pdef}
\end{equation}
with $Z$ a normalization constant and
\begin{equation}
W(\{\lambda_{n}\})=-\sum_{i<j}
\ln|\lambda_{i}-\lambda_{j}|+\sum_{i}V(\lambda_{i}),\label{Wdef}
\end{equation}
\end{mathletters}%
describes an ensemble where {\em all\/} eigenvalue correlations are due to the
jacobian. The potential $V(\lambda)$ determines the mean density
$\rho(\lambda)$ of the eigenvalues, which is non-zero in some interval $(a,b)$.

In many applications of random-matrix theory, it is sufficient to know the
eigenvalue correlations in the bulk of the spectrum, far from the end points at
$a$ and $b$. In some applications, however, the presence of an edge in the
spectrum is an essential part of the problem, and its effect on the spectral
correlations can not be ignored. The application to ``universal conductance
fluctuations'' in mesoscopic conductors is one example
\cite{Sto91,Bee93,Sle93,Bas93}. The application to random surfaces and
two-dimensional quantum gravity is another example \cite{Bre78,Bow91,For93}.

Recently, Br\'{e}zin and Zee \cite{Bre93,Bre93b} reported a remarkably simple
result for the two-level cluster function
\begin{equation}
T_{2}(\lambda,\mu)=-\left\langle\sum_{i\neq
j}\delta(\lambda-\lambda_{i})\delta(\mu-\lambda_{j})
\right\rangle+\rho(\lambda)\rho(\mu),\label{T2def}
\end{equation}
which included the effects of an upper and lower bound on the spectrum. (Here
$\langle\cdots\rangle$ denotes an average with distribution (\ref{PWdef}), and
$\rho(\lambda)=\langle\sum_{i}\delta(\lambda-\lambda_{i})\rangle$ is the mean
eigenvalue density.) For $N\gg 1$, the correlation function (\ref{T2def})
oscillates rapidly on the scale of the spectral band width $(a,b)$. These
oscillations are irrelevant when integrating over the spectrum, so that in the
large-$N$ limit it is sufficient to know the smoothed correlation function.
Br\'{e}zin and Zee considered the unitary ensemble ($\beta=2$), with
$V(\lambda)=\sum_{k=1}^{p}c_{k}\lambda^{2k}$ an even polynomial function of
$\lambda$, so that $a=-b$. (The case $a\neq b$ can then be obtained by
translation of the entire spectrum.) Using the method of orthogonal polynomials
\cite{Meh67}, they computed the smoothed correlation function, with the result
\begin{equation}
T_{2}(\lambda,\mu)=\frac{1}{2\pi^{2}}\,\frac{1}{(\lambda-\mu)^{2}}\,
\frac{a^{2}-\lambda\mu}{[(a^{2}-\lambda^{2})(a^{2}-\mu^{2})]^{1/2}}.
\label{T2BZ}
\end{equation}

The purpose of the present paper is to show how Eq.\ (\ref{T2BZ}) can be
generalized to arbitrary (non-polynomial, non-even) potentials $V(\lambda)$,
and to all three symmetry classes ($\beta=1,2,4$). This universality is
achieved by a functional derivative method \cite{Bee93}, which provides a
powerful alternative to the classical method of orthogonal polynomials. In
Ref.\ \onlinecite{Bee93} we applied this method to the case $a=0$,
$b\rightarrow\infty$ of a {\em single\/} spectral boundary. Here we extend the
analysis to include a finite upper {\em and\/} lower bound on the spectrum.

\section{Method of functional derivatives}
We consider the two-point correlation function
\begin{equation}
K_{2}(\lambda,\mu)=-\left\langle\sum_{i,j}
\delta(\lambda-\lambda_{i})\delta(\mu-\lambda_{j})
\right\rangle+\rho(\lambda)\rho(\mu)\label{K2def}
\end{equation}
(note the unrestricted sum over $i$ and $j$), which is related to the two-level
cluster function (\ref{T2def}) by
\begin{equation}
K_{2}(\lambda,\mu)=T_{2}(\lambda,\mu)-\rho(\lambda)\delta(\lambda-\mu).
\label{KTrelation}
\end{equation}
For $\lambda\neq\mu$ the two correlation functions coincide, so that we can
compare with Eq.\ (\ref{T2BZ}). We prefer to work with $K_{2}$ instead of
$T_{2}$ for a technical reason: Smoothing, in combination with the large-$N$
limit, introduces a spurious non-integrable singularity in $T_{2}(\lambda,\mu)$
at $\lambda=\mu$, while $K_{2}(\lambda,\mu)$ remains integrable.\footnote{The
distinction between $T_{2}(\lambda,\mu)$ and $K_{2}(\lambda,\mu)$ was not made
explicitly in Ref.\ \onlinecite{Bre93}, because only the case $\lambda\neq\mu$
was considered.}

Our analysis is based on the exact relation \cite{Bee93} between the two-point
correlation function $K_{2}(\lambda,\mu)$ and the functional derivative of the
eigenvalue density $\rho(\lambda)$ with respect to the potential $V(\mu)$,
\begin{equation}
K_{2}(\lambda,\mu)=\frac{1}{\beta}\,
\frac{\delta\rho(\lambda)}{\delta V(\mu)}.\label{K2drhodV}
\end{equation}
The smoothed correlator is obtained by evaluating the functional derivative
using the asymptotic ($N\rightarrow\infty$) integral relation between $V$ and
$\rho$,
\begin{equation}
{\cal P}\int_{a}^{b}\!\!d\mu\,\frac{\rho(\mu)}{\lambda-\mu}=
\frac{d}{d\lambda}V(\lambda),\;\;a<\lambda <b.\label{inteq}
\end{equation}
(The symbol ${\cal P}$ denotes the principal value of the integral.)
Corrections to Eq.\ (\ref{inteq}) are smaller by an order $N^{-1}$ for
$\beta=1$ or 4, and by an order $N^{-2}$ for $\beta=2$ \cite{Dys72}. Variation
of Eq.\ (\ref{inteq}) gives
\begin{equation}
\delta b\,\frac{\rho(b)}{\lambda-b}-\delta a\,\frac{\rho(a)}{\lambda-a}+
{\cal P}\int_{a}^{b}\!\!d\mu\,\frac{\delta\rho(\mu)}{\lambda-\mu}=
\frac{d}{d\lambda}\delta V(\lambda),\label{variation}
\end{equation}
with the constraint
\begin{equation}
\int_{a}^{b}\!\!d\lambda\,\delta\rho(\lambda)=0\label{constraint}
\end{equation}
(since the variation of $\rho$ is to be carried out at constant $N$). The end
point $a$ is either a fixed boundary, in which case $\delta a=0$, or a free
boundary, in which case $\rho(a)=0$. Similarly, either $\delta b=0$ or
$\rho(b)=0$. We conclude that we may disregard the first two terms in Eq.\
(\ref{variation}), containing the variation of the end points. What remains is
the singular integral equation
\begin{equation}
{\cal P}\int_{a}^{b}\!\!d\mu\,\frac{\delta\rho(\mu)}{\lambda-\mu}=
\frac{d}{d\lambda}\delta V(\lambda),\;\;a<\lambda <b,\label{variation2}
\end{equation}
which we need to invert in order to obtain the functional derivative
$\delta\rho/\delta V$.

The general solution to Eq.\ (\ref{variation2}) is \cite{Mik64}
\begin{equation}
\delta\rho(\lambda)=\frac{1}{\pi^{2}}\,
\frac{1}{[(\lambda-a)(b-\lambda)]^{1/2}}\left(C-{\cal P}\int_{a}^{b}\!\!d\mu\,
\frac{[(\mu-a)(b-\mu)]^{1/2}}{\lambda-\mu}
\frac{d}{d\mu}\delta V(\mu)\right).\label{generalsol}
\end{equation}
The coefficient $C$ is determined by
\begin{equation}
C=\pi\int_{a}^{b}\!\!d\lambda\,\delta\rho(\lambda).
\label{Csol}
\end{equation}
In view of Eq.\ (\ref{constraint}), we have $C=0$. Combination of Eqs.\
(\ref{K2drhodV}) and (\ref{generalsol}) yields the two-point correlation
function
\begin{equation}
K_{2}(\lambda,\mu)=\frac{1}{\beta\pi^{2}}\,\frac{1}
{[(\lambda-a)(b-\lambda)]^{1/2}}\,
\frac{\partial}{\partial\lambda}\frac{\partial}{\partial\mu}
\left(
[(\mu-a)(b-\mu)]^{1/2}\ln|\lambda-\mu|\right).
\label{K2result}
\end{equation}
(We have substituted the representation ${\cal P}x^{-1}=(d/dx)\ln|x|$ for the
principal value.)

The two-point correlation function (\ref{K2result}) has an {\em integrable\/}
singularity for $\lambda=\mu$. For $\lambda\neq\mu$ one can carry out the
differentiations, with the result
\begin{equation}
K_{2}(\lambda,\mu)=\frac{1}{\beta\pi^{2}}\,\frac{1}{(\lambda-\mu)^{2}}\,
\frac{\frac{1}{2}(a+b)(\lambda+\mu)-ab-\lambda\mu}
{[(\lambda-a)(b-\lambda)(\mu-a)(b-\mu)]^{1/2}}
\;\;{\rm if}\;\lambda\neq\mu.\label{T2result}
\end{equation}
For $a=-b$ and $\beta=2$ we recover the formula (\ref{T2BZ}) of Ref.\
\onlinecite{Bre93} for even polynomial potentials in the unitary ensemble.  For
$a=0$ and $b\rightarrow\infty$ we recover the correlator of Ref.\
\onlinecite{Bee93},
\begin{eqnarray}
K_{2}(\lambda,\mu)&=&\frac{1}{\beta\pi^{2}}\,
\frac{\partial}{\partial\lambda}\frac{\partial}{\partial\mu}
\ln\left|\frac{\surd\lambda-\surd\mu}
{\surd\lambda+\surd\mu}\right|\nonumber\\
&=&\frac{1}{2\beta\pi^{2}}\,(\lambda-\mu)^{-2}
(\lambda+\mu)(\lambda\mu)^{-1/2}
\;\;{\rm if}\; \lambda\neq\mu,
\label{T2single}
\end{eqnarray}
for the case of a single spectral edge.

An important application of the smoothed two-point correlation function is to
compute the large-$N$ limit of the variance ${\rm Var}\,A\equiv\langle
A^{2}\rangle-\langle A\rangle^{2}$ of a linear statistic
$A=\sum_{n=1}^{N}a(\lambda_{n})$ on the eigenvalues, by means of the
relationship
\begin{equation}
{\rm Var}\,A=-\int_{a}^{b}\!\!d\lambda\int_{a}^{b}\!\!d\mu\,a(\lambda)
a(\mu)K_{2}(\lambda,\mu).\label{VarAdef}
\end{equation}
Substituting Eq.\ (\ref{K2result}) we obtain, upon partial integration,
\begin{equation}
{\rm Var}\,A=\frac{1}{\beta\pi^{2}}\,{\cal P}
\int_{a}^{b}\!\!d\lambda\int_{a}^{b}\!\!d\mu\,
\left[\frac{(\mu-a)(b-\mu)}{(\lambda-a)(b-\lambda)}\right]^{1/2}\,
\frac{a(\lambda)}{\lambda-\mu}\,\frac{d}{d\mu}a(\mu).
\label{VarAresult}
\end{equation}
Note that here it is essential to work with the expression (\ref{K2result}) for
$K_{2}(\lambda,\mu)$, which is integrable, and that one can not use the
expression (\ref{T2result}), which has a spurious non-integrable singularity at
$\lambda=\mu$. The formula (\ref{VarAresult}) is the generalization to a
spectrum bounded from above and below of previous formulas by Dyson and Mehta
\cite{Dys63} (for an unbounded spectrum) and by the author \cite{Bee93} (for  a
spectrum bounded from below).

\section{Conclusion}

The result (\ref{K2result}) for the smoothed two-point correlation function in
the large-$N$ limit holds for {\em all\/} random-matrix ensembles of
Wigner-Dyson type, i.e.\ with a probability distribution of the general form
(\ref{PWdef}). The form of the eigenvalue potential $V(\lambda)$ is irrelevant.
It is also irrelevant whether the end point at $\lambda=a$ (or at $b$) is a
fixed or a free boundary. This is remarkable, because the eigenvalue density
behaves entirely different in the two cases: At a fixed boundary
$\rho(\lambda)$ {\em diverges\/} as $(\lambda-a)^{-1/2}$, while at a free
boundary $\rho(\lambda)$ {\em vanishes\/} as $(\lambda-a)^{1/2}$. Both cases
are of interest for applications: The spectrum considered in Ref.\
\onlinecite{Bre93}, in connection with two-dimensional gravity, has free
boundaries; The spectrum considered in Ref.\ \onlinecite{Bee93}, in connection
with mesoscopic conductors, has a fixed boundary.

While the form of the eigenvalue potential is irrelevant, the form of the
eigenvalue interaction does matter. Consider an eigenvalue distribution
function of the form (\ref{PWdef}), but with a {\em non-logarithmic\/}
eigenvalue interaction $u(\lambda,\mu)\neq\ln|\lambda-\mu|$. Such a
distribution describes the energy level statistics of disordered metal
particles \cite{Jal93}, and the statistics of transmission eigenvalues in
disordered metal wires \cite{Bee94}. The analysis of Sec.\ 2 carries over to
this case, but the integral kernel $(\lambda-\mu)^{-1}$ in Eqs.\ (\ref{inteq})
and (\ref{variation2}) has to be replaced by the kernel $\partial
u/\partial\lambda$. The two-point correlation function now equals $1/\beta$
times the inverse of this integral kernel, and differs from the result
(\ref{K2result}) for a logarithmic interaction.

So far we have only considered the two-point correlation function
$K_{2}=\beta^{-1}\delta\rho/\delta V$ and the closely related two-level cluster
function $T_{2}$. Br\'{e}zin and Zee \cite{Bre93} also computed the three- and
four-level cluster functions, and found that they vanish identically upon
smoothing. The linearity of the relation (\ref{inteq}) between $\rho$ and $V$
implies in fact that this holds for {\em all\/} higher-order cluster functions.
This argument is equivalent to Politzer's proof \cite{Pol89} that any linear
statistic on the eigenvalues has a gaussian distribution in the large-$N$
limit.

\acknowledgments
A valuable discussion with E. Br\'{e}zin is gratefully acknowledged. This
research was supported in part by the ``Ne\-der\-land\-se or\-ga\-ni\-sa\-tie
voor We\-ten\-schap\-pe\-lijk On\-der\-zoek'' (NWO) and by the ``Stich\-ting
voor Fun\-da\-men\-teel On\-der\-zoek der Ma\-te\-rie'' (FOM).

\end{document}